\documentclass[epj]{svjour}

\usepackage{graphicx}
\graphicspath{{Figure/}}

\def\unit#1{\ensuremath{\mathrm{~#1}}}
\def\unite#1{\ensuremath{\mathrm{~#1}}}
\def\ket#1{\ensuremath{\left|#1\right>}}
\def\TRamsey{\ensuremath{{T_{\mathrm{Ramsey}}}}}

\def\Tacc{\ensuremath{T_\mathrm{acc}}}
\def\deltasep{\ensuremath{\delta_\mathrm{sep}}}
\def\braket#1#2{{\ensuremath{\left<{#1}\left|\vphantom{#1}%
      \vphantom{#2}\right.{#2}\right>}}}
\def\ketbra#1#2{\ensuremath{\left|#1\vphantom{#2}\right>
    \left<#2\vphantom{#1}\right|}}

\def\poub#1{}
\def\etal{\textit{et al.}}

\def\vert#1{{\color{green}}}
\usepackage{amsmath}
\usepackage{amsfonts}
\begin{document}

\title{Theoretical Analysis of a Large Momentum Beamsplitter using Bloch Oscillations}

\author{Pierre~Clad\'e, Thomas~Plisson, Sa\"\i da~Guellati-Kh\'elifa, Fran\c cois~Nez, Fran\c cois~Biraben}
\institute{Laboratoire Kastler Brossel, UPMC, \'Ecole Normale
Sup\'erieure, CNRS,  4 place Jussieu, 75252 Paris Cedex 05, France}

\abstract{In this paper, we present the implementation of Bloch oscillations in an atomic interferometer to increase the separation
of the two interfering paths.
A numerical model, in very good agreement with the experiment, is developed. The contrast 
of the interferometer and its sensitivity to phase fluctuations and to intensity fluctuations are also calculated. We demonstrate that the 
sensitivity to phase fluctuations can be significantly reduced by using a suitable arrangement of Bloch oscillations pulses.
}

\authorrunning{Clad\'e \etal}
\titlerunning{Large Momentum Beamsplitter using Bloch Oscillations}

\maketitle

%
%
%

\maketitle

\section{Introduction}

Atom interferometry is nowadays a corner stone in high precision
measurement. It has been used to determine the fine structure constant
\cite{Wicht:02,Cadoret2008}, as well as the Newton's gravitational
constant \cite{Fixler2007,lamporesi_PRL_2008}, to test general
relativity \cite{muller_2008_PRL031101} and to measure gravity
\cite{Peters1999,LeGouet:2007}.  In these applications, the
interferometers rely on momentum exchange between light and atoms to
split and recombine atomic wave-packets.  The sensitivity of the
measurements is proportional to the spatial separation of the
wave-packets in the interferometer.  As the beamsplitters are performed
with two photon Raman transition, the velocity difference between the
arms of the interferometer $\delta v $ is equal to $2v_r$ (where $v_r
= \hbar k/m$ is the recoil velocity, $k$ is the wavenumber of light
and $m$ the mass of the atoms). The sensitivity of the interferometer
is thus proportional to $2v_r T$ where $T$ is the typical duration of
the interferometer\footnote{For a gravimeter, the velocity change due to
gravity is also proportional to $T$, which leads to a sensitivity 
to gravity scaling as $T^2$}.

Until recently, the main way to improve the sensitivity of
interferometers in order to achieve higher precision measurements was
to increase the duration $T$. However, it requires a careful control
of phase noise of the lasers as well as vibrations in order to keep a
good signal to noise ratio. Transversal spread of the atomic cloud or
motion of the cloud inside the science chamber set a limit to this
interaction time. Interaction times of the order of several hundreds
of\unit{ms} are achieved \cite{Peters1999}.

Another method to increase the separation of the wave\-packets and so
the sensitivity of the interferometer consists in using a large
momentum transfer pulse with a separation $\delta v > 2v_r$. High order Bragg
diffraction of matter wave can be used to increase $\delta v$
\cite{muller:180405}. However, the laser power required for Bragg
diffraction increases sharply with $\delta v$
\cite{muller2009_PRL240403}. Recently, a double-diffraction technique
has been proposed to enhance the area of a Raman atom interferometer
\cite{leveque:2009}. Increasing $\delta v$ can be also done with Bloch
oscillations. It has been suggested and demonstrated by the group of
W.D.~Phillips on a Bose Einstein condensate \cite{Denschlag}. The
coherence of the acceleration with Bloch oscillations (BO) is a key
point of the process \cite{Peik}.  More recently, a group at
Stanford/Berkeley and our group have implemented this method by
inserting the so called large momentum transfer beam splitter (LMTBS)
in an atom interferometer \cite{clade_PRL2009,muller2009_PRL240403}.

The large momentum transfer beam splitter (LMTBS) is realized with a
suitable combination of two processes.  The first one, realized with
usual methods (Bragg pulse in the group of Stanford/Berkeley and Raman
pulses in ours), implements a separation between two wavepackets. The
second process is used to selectively and coherently accelerated only
one of the wavepackets with Bloch oscillations in order to increase
$\delta v$. This separation keeps the coherence between the two wave
packets.

Our experiment where we demonstrate the realisation of a LMTBS has been
described in a letter \cite{clade_PRL2009}. In this paper, we present a
model that help us to calculate precisely the efficiency of the LMTBS
(first part) and also the phase of the interferometer (second part). The calculation
of this phase is a key point to understand the sensitivity of the interferometer (third part).


\section{Bloch oscillations}

In this section we describe the Bloch oscillation process that
is used to increase the separation of atoms in the interferometer. 
The phenomenon of Bloch oscillation is a well understand tool 
to coherently accelerate ultra-cold atoms\cite{BenDahan}. 
In order to realise a LMTBS, the difficulty is not only to 
efficiently accelerate atoms but to be able to accelerate
only atoms from on arm of the interferometer, keeping the coherence 
between the two arms.


\subsection{Phenomenological description of the acceleration}

\begin{figure}
\begin{center}
\includegraphics[width = .9\linewidth]{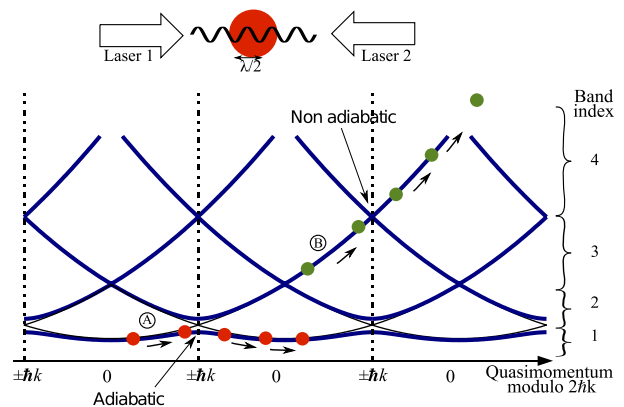}
\end{center}
\caption{Band structure of the optical lattice. This graph represents the band energy in the frame of the lattice as a function
of the quasi-momentum. Because quasi-momentum is defined modulo $2\hbar k$, one can think of this graph as an unrolled cylinder
where the vertical dot-dashed lines are overlapping.}
\label{fig:bande}
\end{figure}

The Bloch oscillation acceleration is based on coherent
transfer of momentum between an accelerated optical lattice and atoms. 
The optical lattice results from interference of the two lasers beams of the same intensity $I$, detuned
by $\Delta$ from the one photon transition. The amplitude of the resulting standing wave is $4I$. Atoms endure
a periodic potential $U(x)$ due to light-shift (a.c. stark effect) which can be expressed as:
\begin{equation}
U(x) = \frac{U_0}2 \left(1+\cos(2kx)\right),
\label{eq:Potential}
\end{equation}
where $k=2\pi/\lambda = 2\pi \nu /c$ is the wavevector of the laser,
$U_0 = h\frac{\Gamma^2}{8\Delta}\frac I{I_S}$ with $\Gamma$
representing the natural linewidth of the excited state and $I_S$ its
saturation intensity. 
In the
following, the constant part $U_0/2$ which does not play any role for
the explanations is removed from every equation.  Because of the
periodicity of the potential, the eigenenergies of the system show a
band structure \cite{Bloch}.  Each eigenstate is described by a band
index $n$ and a quasimomentum $q$ defined modulo $2\hbar k$ \cite{Ashcroft_en}.  The
Fig.~\ref{fig:bande} represents this band structure, where the first
Brillouin zone is unfolded. The splittings between bands at the center
($q=0$) and at the edge ($q=\hbar k$) of the first Brillouin zone
occurs where atoms are resonantly coupled by $U(x)$. 

In a quantum optics
view, the avoided crossing (Bragg resonance) between the first and
second bands is connected to the resonance of the two photon
transition between states of momentum $p=\pm \hbar k$. Similarly, the
splitting between the second and third bands is due to a resonant four photon transition between
states of momentum $p=\pm 2\hbar k$, and so on for the $n^\mathrm{th}$
and $(n+1)^\mathrm{th}$ band. The order of the transition grows with
the band index, thus the higher the band index is, the weaker is the
splitting between the bands. 

When the atoms in the first
band are subjected to a constant and uniform force, their
quasimomentum increases linearly with time. If the force is weak
enough to induce adiabatic transition, the atoms stay in the first
band and therefore have a periodic motion (Bloch oscillations
\cite{Zener}). The time necessary for a variation of the quasimomentum
of $2\hbar k$ is the period of the oscillations.  
For atoms in higher band, for a constant acceleration, the probability to make an adiabatic
transition (stay in the same band) will be higher for low values of
the band index.

Bloch oscillations were initially introduced in the case of a lattice and a uniform force. In our
experiment, a
constant acceleration is applied to the lattice. 
This acceleration acts like a force in the frame of the lattice (see Ref.~\cite{Peik}), 
and an equivalent of Bloch oscillations can be observed \cite{BenDahan}: the 
adiabatic transition corresponds to
a change of velocity of 2 recoils, whereas a non-adiabatic transition does not
change the velocity of the atom (free particle). 

In our LMTBS, one wavepacket is loaded in the first band (A, see
Fig.~\ref{fig:bande}) and the second in the third band (B).  
The acceleration is small enough so that atoms
in the first band are likely to make an adiabatic transition, but
strong enough so that atoms initially in the third band change
band. The atoms in the first band are periodically accelerated by 2 recoils momentum. 
In the same time, atoms in the third band are not
accelerated.  This process allows us to selectively accelerate atoms
depending on their initial velocity.

Note that in our experiment, atoms with different velocities have also
different internal states because of the Raman transition used to
create the initial splitting. The amplitude of the lattice is then
different for the two wavepackets.  This effect could be used to
select the wavepacket which is going to be accelerated. However for
our experimental parameters, the difference in light shift is
negligible and so the selection of the accelerated wavepacket relies
only on its initial velocity and not its internal state.

\subsection{Theoretical description of LMTBS}

In order to optimize the efficiency of the LMT beamsplitter, the evolution of the atomic system
has been computed. In the laboratory frame, the Hamiltonian of the particle is:
\begin{equation}
H = \frac{\hat p^2}{2m} + V(\hat x,t)
\label{eq:Hamilt1}
\end{equation}
where $V(x,t)$ is the accelerated periodic potential. If we denote by $U_0(t)$  the amplitude 
of the lattice and by $\delta \nu(t)$ 
the frequency difference between the two laser beam. Then $v(t) = 2k\delta\nu(t)$ is the velocity of the lattice and 
\begin{equation}
V(x,t) =  \frac{U_0(t)}2 \cos\left(2k\left(x - \int^t v(t^\prime)\mathrm dt^\prime \right)\right)
\label{eq:V}
\end{equation}

We should note that, even if the lattice is accelerated, in the laboratory frame, this 
Hamiltonian is still periodic in space. Therefore the quasimomentum is still a good
quantum number and is conserved. This Hamiltonian cannot be solved efficiently, because of 
the fast time variation of the potential. One could 
write the Hamiltonian in the accelerated frame. This Hamiltonian will be time independent, 
but will no longer be periodic, and the quasi-momentum will not be conserved. 

We use another and convenient transformation which consists in translating the wave function in position space 
by the quantity $X = \int^t v(t^\prime)\mathrm dt^\prime$:
\begin{eqnarray}
x^\prime = x + X(t) \\
\psi^\prime(x^\prime) = \psi(x)
\end{eqnarray}
The Hamiltonian becomes:
\begin{equation}
H(t) = \frac{\left(\hat{p}-mv(t)\right)^2}{2m} + \frac{U_0(t)}2\cos\left(2k\hat{x}\right),
\label{eq:Hamilt2}
\end{equation}
where the energy shift $mv(t)^2/2$ is left out. This Hamiltonian described the Bloch oscillation in the "solid-state physics" point of view \cite{Peik}. The advantage of this presentation is that the Hamiltonian is time independent when the lattice is moving at a constant velocity. 

If we start with a plane wave of momentum $p_0$ (and therefore of
quasi-momentum $q_0 = p_0 \mathrm{\ modulo\ } 2\hbar k)$), the
wavefunction will remain in the subspace of states of quasi-momentum
$q_0$. We therefore choose to solve the problem in the basis of the
plane wave of momentum $p_0 + 2l\hbar k$, $l\in\mathbb{Z}$ which is a
basis of this subspace. 


To solve numerically the problem i npractice, the momentum
basis is restricted to a finite number of $l$, $|l| \leq
N_\mathrm{cut}$. The value of $N_\mathrm{cut}$ is set by the maximal velocity
of atoms in the optical lattice. It results from the sum of the
number of oscillations (velocity of the atom) and the width of the velocity distribution of the
Bloch states ($\Delta p / \hbar k $). This width
$\Delta p$ is set by the Wannier function in momentum space\cite{clade:052109}
which is given by the wave function of an atom trapped
in a single well without tunnelling between sites. The frequency
$\omega$ of the trap is proportional to $\sqrt{\left|U_0\right| E_r}$.
For the ground state of an harmonic oscillator, $\Delta p =
\sqrt{\hbar m \omega /2}$. We obtain that $\Delta p / \hbar k \simeq
\sqrt[4]{U_0 / E_r}$ \cite{clade:052109,Hartmann_NJP}. In our calculation, we have
used $N_\mathrm{cut} = 10$ for 2 oscillations and $U_0 \simeq 8E_r$ .

The Hamiltonian is calculated using the dimensionless parameters
$\kappa = U_0 /8E_r$ (where $E_r = \frac{\hbar^2k^2}{2m}$ is the
recoil energy) and $p(t)=(p_0 + mv(t))/\hbar k$. All the units are
scaled so that $\hbar = k = m = 1$.

The expression of the Hamiltonian $H$ and the ket \ket{\psi} are reduced to :
\begin{equation}
H = \begin{bmatrix}
    \ddots &    \kappa &    0       &    \ldots       & 0    & \ldots& 0 \\
    \kappa &  p_{-l}^2/2 & \kappa     & \ddots        &    \vdots   &    & \vdots     \\
    0       &   \kappa &   \ddots &    \kappa  &  0&  \ldots &   0 \\
     \vdots   &   \ddots  &   \kappa  &   p_0^2/2 &   \kappa  &   \ddots  & \vdots\\
     0 &   \ldots    &       0   &   \kappa& \ddots&\kappa& 0  \\
     \vdots       &       &    \vdots       & \ddots  &\kappa & p_l^2/2 & \kappa \\
    0       &    \ldots   &   0 &      \ldots     &    0  & \kappa & \ddots \\
\end{bmatrix},\,
\ket{\psi} = \begin{bmatrix}
	\vdots \\
	c_{-l} \\
	\vdots \\
	c_0 \\
	\vdots	\\
	c_{l} \\
	\vdots	\\
\end{bmatrix}
\label{eq:supermatrice}
\end{equation}
where $p_l = p(t) + 2l$. Each coefficient $c_l$ represents
the amplitude of probability for the state to be in the plane wave of momentum $2l\hbar k$.

The LMT pulse is switched on (and off) adiabatically to maximize the
number of atoms transferred from an atomic state with well known
momentum $p_0$ (plane wave) to a Bloch state with a quasimomentum $q_0
= p_0$ in the band $n$ (and reciprocally)\cite{clade:052109}.  In
between, the lattice is accelerated to transfer a given number of
recoils to one component of the wavefunction. We denote by
$t_{\mathrm{adiab}}$ the duration of the linear ramp used to switch on
and off the lattice, $T_\mathrm{acc}$ the duration of the
acceleration, $2N$ the number of transferred recoils, $U_0$ the
maximal amplitude of the lattice and $p_0$ the initial momentum of the
atoms.  

\subsection{Optimization of the acceleration}
\begin{figure}
\begin{center}
\includegraphics[width = .9\linewidth]{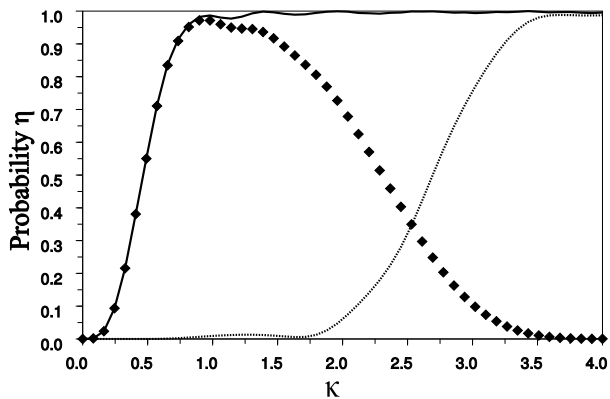}
\end{center}
\caption{Transfer probability as a function of the maximal optical depth $\kappa = U_0/8E_r$ of the lattice. The acceleration is in 200\unit{\mu s} for N=2 recoils. Solid line: transfer probability for the first band $\eta_{11}$; dashed line: for the third band, $\eta_{33} \approx 1-\eta_{34}$; diamond: Efficiency of the LMT pulse, $\eta= \eta_{11}\eta_{34}$. }
\label{fig:probaU0}
\end{figure}

\begin{figure}
\begin{center}
\includegraphics[width = .9\linewidth]{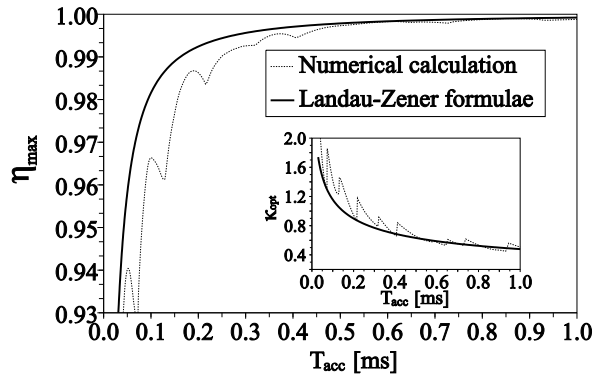}
\end{center}
\caption{Efficiency of the LMT pulse as a function of the total duration of the acceleration for two Bloch oscillations. For each value of the duration, we calculate the value $\kappa_\mathrm{opt}$ of the amplitude $\kappa$ of the lattice that optimizes the efficiency. This value is plotted in the inset of the figure.
The "saw tooth" shape of $\kappa_\mathrm{opt}$ (and therefore of $\eta_\mathrm{opt}$) is due to small oscillations between the bands. They do not appear in the Landau-Zener model which doesn't take into this effect (lost atoms cannot go back into their initial band).}
\label{fig:etavsT}
\end{figure}

The optimization of the acceleration results from a compromise
between an adiabatic acceleration of the atoms in the first band and a
strong enough (non-adiabatic) acceleration of the atoms in the third
band.  Reciprocally, for a given acceleration, the amplitude of the
lattice should be in a given range in order to satisfy this latest
compromise. In our experiment, it is easier to vary the amplitude of
the lattice. Therefore, we have plotted on the Fig.~\ref{fig:probaU0}
the probability for an atom to stay in its band as a function of the
lattice amplitude $\kappa$ for the first (solid line) and third
(dashed line) bands.  There is clearly an intermediate regime where
the probability $\eta_{11}$ for an atom to stay in the first band is
high whereas the probability $\eta_{34}$ for an atom to leave the
third band (and reach the fourth one) is also high.  The efficiency of
the LMT is calculated as
\begin{equation}
\eta = \eta_{11}\eta_{34}.
\end{equation}
It sets the contrast of the interference pattern of the two wavepackets. The value of $\eta$
is plotted on Fig.~\ref{fig:probaU0}.  For our experimental parameters
(an acceleration of 2 recoils in $200\unit{\mu s}$), the maximum
efficiency is about $98\%$ for $\kappa = 1$.  This maximum
$\eta_\mathrm{max}$, and the optimal value $\kappa_\mathrm{opt}$ of
$\kappa$ depends upon \Tacc. The values of $\eta_\mathrm{max}$ and
$\kappa_\mathrm{opt}$ versus $\Tacc$ are plotted on
Fig.~\ref{fig:etavsT}. For an infinite duration of $\Tacc$, the
efficiency is closed to one while the value for $\kappa_\mathrm{opt}$
is nearby zero.

This behavior is understandable using the Landau-Zener criteria which can be used to calculate transition probabilities
in the weak binding limit ($\kappa \ll 1$), \cite{Peik}:
\begin{eqnarray}
\eta_{11} & =& 1- \exp\left(-\pi \frac{\kappa^2}{a}\right) \\
\eta_{34} & =&  \exp\left(-\pi \frac{\kappa^6}{768 a}\right)
\label{eq:LZ}
\end{eqnarray}
where $a = m^2/\hbar^2k^3\,\mathrm dv/\mathrm dt $, is the dimension-less acceleration of the lattice. We clearly see that
for a small value of $\kappa$, there is a value of $a$ so that $\frac{\kappa^6}{768 a} \ll 1$ and
$\frac{\kappa^2}{a} \gg 1 $ leading to efficiencies $\eta_{11}$ and $\eta_{34}$ close to one. In fact, there is no
theoretical limitation to the weakness of $\kappa$. There is always a suitable value of
$a$ which sets $\eta$ close to unity. However, this suitable value of $a$ becomes very small, which means
that an infinite amount of time
is required for the acceleration. In an other way, with a given value of $a$, the higher
is $a$, the lower is $\eta$ with an optimized $\kappa$.

In the experimental point of view, the duration of LMT pulse can not be extend as much as we want. This duration has to be smaller
than the time separation between the $\pi/2$ pulses of our interferometer. The time separation between $\pi/2$ pulses results
from a compromise between the narrowest width of the atomic fringes (and so the resolution of the interferometer)
and the vibrations of the experimental set up.
Practically, the comparison of the theoretical and experimental results presented in this paper is done with $\Tacc=0.2\unit{ms}$
(for two oscillations). We will briefly explain at the end of this paper
a way to transfer a larger number of recoils using a non constant acceleration. 


\subsection{Adiabatic loading of atoms in the lattice}
In the previous section, the efficiency $\eta$ of the separation has been calculated assuming that
atoms are initially in a well defined Bloch state. Yet, in our experiment, we deal
with a subrecoil velocity distribution, thus the best description of the initial state is in a plane wave basis.
To load the atoms into a Bloch state, the lattice depth is increased adiabatically. Atoms in an initial momentum $p_0$
are loaded into the Bloch state \ket{n,p_0}, where the band index $n$ is the nearest integer above $\left| p_0/\hbar k \right|$.

\begin{figure}
\begin{center}
\includegraphics[width = .9\linewidth]{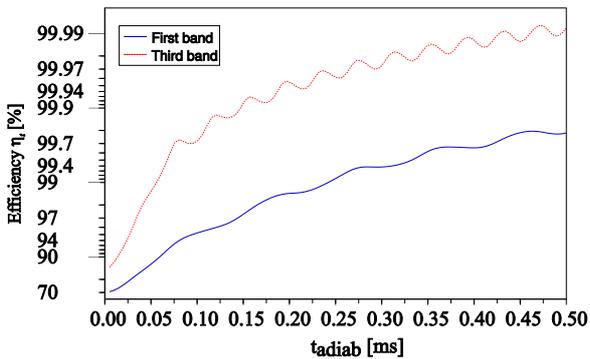}
\end{center}
\caption{\label{fig:Eff_vs_Tadiab}Transfer probability for the loading of atoms in the optical lattice as a function of loading time $t_\mathrm{adiab}$. The quasi-momentum is $q_0 = 0.5$ and the final amplitude of the lattice is $U_0 = 8E_r$.}
\end{figure}

The amplitude of probability $\eta_{t}$ to load atoms in a given band is defined as
$\left| \braket{\psi}{n,p_0}\right|$ where $\ket \psi$ is the wave function and \ket{n,p_0} the Bloch wave function.
This probability depends upon the loading time $t_\mathrm{adiab}$, the final depth of the lattice $U_0$ (the amplitude of the
lattice is linearly increased) and the momentum $p_0$. The
Fig.~\ref{fig:Eff_vs_Tadiab} shows, as functions of the duration $t_\mathrm{adiab}$, the probabilities $\eta_{t}$ to load atoms
either in the first or in the third band.
With an initial quasi-momentum $q_0 = 0.5$, the probability $\eta_{t}$ is higher for the third band
than for the first one because the coupling with the lattice gets weaker with the increase of band index.

\begin{figure}
\begin{center}
\includegraphics[width = .9\linewidth]{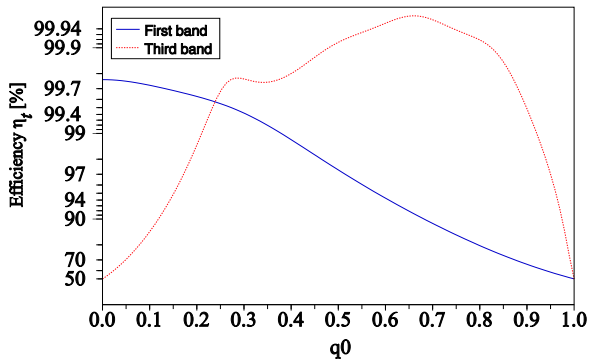}
\end{center}
\caption{Transfer probability for the loading of atoms in the optical lattice as a function of the initial quasi-momentum $q_0$ in unit of $\hbar k$ for atoms in the first and third band of the lattice, $t_{\mathrm{adiab}}=150\unit{\mu s}$ and $\kappa = 1$.}
\label{fig:eff_vs_q0}
\end{figure}

On Fig.~\ref{fig:eff_vs_q0}, this efficiency $\eta_{t}$ is plotted versus the initial quasi-momentum for
$t_\mathrm{adiab} = 150 \unite{\mu s}$ and for $U_0 = 8 E_r$.

In the case of atoms loaded in the first band, the efficiency is maximum at the center of the first Brillouin zone ($p_0 = 0$).
At the edge of the first Brillouin zone ($p_0 = \hbar k$), $\eta_{t}$ is exactly 50\%. In this situation, it is indeed
not possible to adiabatically transfer the atoms into the optical lattice, because the initial state is degenerate.
The same phenomena occurs on the two edges of the third zone ($p_0 = 2\hbar k$ and $p_0 = 3\hbar k$). The same difficulty
happens at the end of the LMT pulse when the atoms are transferred back from the lattice to a well defined momentum state
(plane wave).

\subsection{Total efficiency of the LMT}

\begin{figure}
\begin{center}
\includegraphics[width = .7\linewidth]{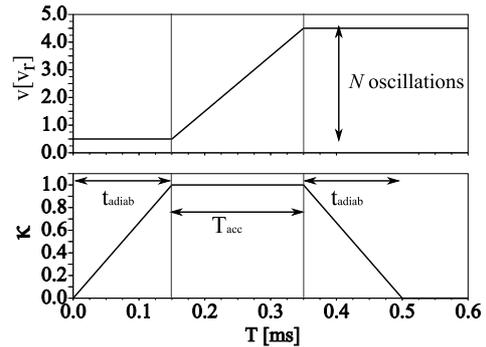}
\end{center}
\caption{Time evolution of the amplitude $\kappa = U_0/8E_r$ of the lattice and of the velocity $v$ of the lattice for a beam splitter with 2 oscillations. These parameters correspond to an acceleration of about 100\unit{m.s^{-1}}.}
\label{fig:2osc}
\end{figure}

We have calculated the total efficiency $\eta_\mathrm{tot}$  with the
following parameters: $t_{\mathrm{adiab}}=150\unit{\mu s}$, $\Tacc=200\unit{\mu s}$,
$N=2$, $U_0 = 8 E_r$. Those parameters
correspond to an optimization of the efficiency, keeping the total time (defined as $2t_{\mathrm{adiab}} + \Tacc$) constant but varying the amplitude of the lattice and
$t_{\mathrm{adiab}}$ (see Fig.~\ref{fig:2osc}). 

\begin{figure}
\begin{center}
\includegraphics[width = .9\linewidth]{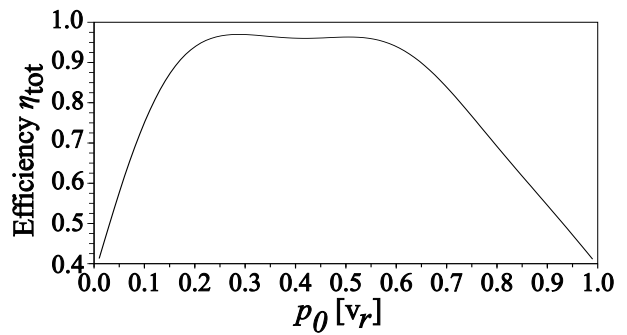}
\end{center}
\caption{Total efficiency $\eta_\mathrm{tot}$ of the LMT pulse as a function of the initial momentum $p_0$.}
\label{fig:EffFonctQ}
\end{figure}

The total efficiency is plotted as a function of the initial momentum $p_0$ on Fig.~\ref{fig:EffFonctQ}. It is computed including the contributions of the adiabatic loading (and unloading) of the atoms
in the lattice. 
In the first band, the state of initial quasimomentum $q_0 = 1$ and in the third bands, the states of initial quasi-momentum $q_0=0$ and $q_0=1$ are degenerated states. Therefore these atoms cannot be adiabatically loaded in the lattice. 
Consequently, it is important
to optimize the width of the initial momentum distribution such that it fits in the range where the process is very efficient.
However, the wider is
the initial velocity distribution loaded into the LMT pulse, the higher is the number of atoms that contributes to the interferometer
and so the larger is the signal to noise ratio. Nevertheless, note that the total efficiency is larger than 95\% on a wide range
of $p_0$.


\section{Phase shift of the interferometer}

In the preceding section, we have calculated the efficiency of the LMT beamsplitter. This efficiency 
is one of the key parameter for calculating the contrast of the interferometer. Another parameter
of the interferometer, is the phase shift. In the following section, we describe in detail how it can be calculated. 

\subsection{Ramsey-Bord\'e interferometer}
\label{section2}

\begin{figure}
\begin{center}
\includegraphics[width = .7\linewidth]{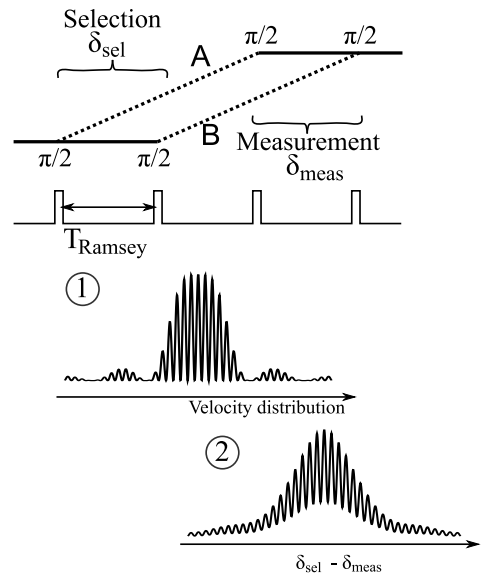}
\end{center}
\caption{Scheme of the usual Ramsey Bord\'e interferometer. A first
  set of two $\pi/2$ pulses, separated by a delay $\TRamsey$, select a
  Ramsey fringe pattern in velocity space (1). A second set of two
  $\pi/2$ pulses, with the same delay is used to close the
  interferometer. The probability for an atom to be transferred by the
  second set of $\pi/2$ pulses from one state to the other is recorded
  as a function of the difference between the frequency of the laser
  during selection $\delta_\mathrm{sel}$ and measurement
  $\delta_\mathrm{meas}$ (2). This interferometer measure then the
  velocity variation between the selection and measurement pulses.}
\label{fig:RBSimple}
\end{figure}

Let us briefly recall the principle of a Ramsey Bord\'e interferometer
realized with Raman pulses (see Fig.~\ref{fig:RBSimple}). It is based
on a succession of four $\pi/2$ Raman pulses (or, a succession
$\pi/2$-$\pi$-$\pi/2$). 
Each Raman pulse couples two states (labeled $\ket{g_1}$ and
$\ket{g_2}$) of the hyperfine structure of the ground state of the
atom via a two photon transition in a $\Lambda$ scheme. The two laser
used for the transition are counterpropagating, therefore, the two
states have different velocities. This allow us to split the initial wave
packet in two. The relative phase between the two states is determined by
the phase of the laser beams used for the Raman transition.
 The second and third pulse are used as mirror to
modify the trajectory of atoms and the fourth pulse is used to recombine the two
wavepackets. 

A simple way to precisely understand the output fringes pattern of the
interferometer is to consider the contribution of each pair of $\pi/2$
pulses. After the first pair of $\pi/2$-pulses, the velocity
distribution of atoms in state \ket{g_2} follows a Ramsey fringes
pattern (see Fig.~\ref{fig:RBSimple}).  Thanks to the Doppler effect,
the frequency difference between Raman beams sets the center of the
Ramsey fringes pattern and select the most populated velocity class in
the initial velocity distribution. Atoms which have not been
transferred are eliminated.  The second set of $\pi/2$ pulses, also
Doppler sensitive, is used to probe the selected velocity
distribution.  The resulting signal of the interferometer is the
product of these two Ramsey distributions. By scanning the frequency
difference of the lasers used for the second set of pulses, the second
Ramsey pattern is translated in velocity space. The resulting signal
versus laser frequency difference is the convolution of the two
distributions created with each pair of $\pi/2$ pulses. The center of
this convolution profile gives access to the velocity of the selected
atoms. 

The resolution of the velocity measurement is given by the fringe width.
For a counterpropagating Raman transition, the Doppler shift
of an atom at velocity $v$ is $2kv$. The frequency splitting of the
fringes in the convolution profile is $2\pi/\TRamsey$, \textit{i.e.}
$\lambda/2\TRamsey$ in velocity space. This expression can also be
understood as resulting from the interference of two selected wave
packets: they are spatially separated by $\Delta X = 2v_r
\TRamsey$. In velocity space, the interference produces fringes with a
separation $\frac h{m\Delta X} = \frac \lambda{2\TRamsey}$.

\begin{figure}
\begin{center}
\includegraphics[width = .8\linewidth]{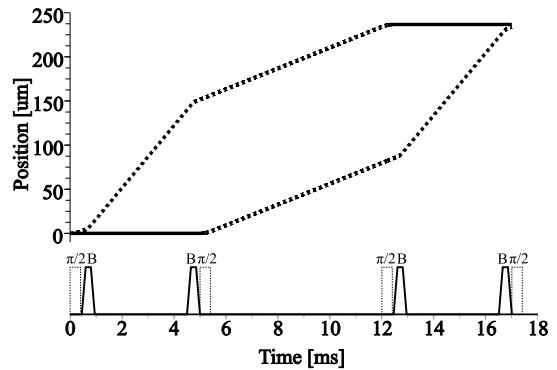}
\end{center}
\caption{Scheme of an interferometer using LMT beamsplitters. With
  this scheme four LMT Bloch pulses (labelled B) are inserted between
  the Raman $\pi/2$ pulses. In the experiment, we use a more
  sophisticated scheme described Fig.~\ref{fig:sequence}.}
\label{fig:LMTSimple}
\end{figure}

There are different ways of modelling an atomic interferometer. One can, for example, use
the Feynman path integral method \cite{Borde1989,Berman}. However, this method
does not fit very well with the description of Bloch oscillation
made in the first section of this paper, where we use plane wave. 

We therefore describe the atom interferometer
using also a plane wave basis. We will first introduce how 
this description can be used for a regular Ramsey-Bord\'e interferometer
and then extend it to the LMT interferometer.

\subsection{Modelisation of the interferometer}
\def\VR{V_{\mathrm R}}
\def\keff{\ensuremath{k_\mathrm{eff}}}

We are going to calculate the evolution of a wave packet 
through the interferometer. This wavepacket will be descibed
as a superposition of plane waves with a mean momentum $p_0$. During the 
interferometer the packet will be split into two paths (labeled
A and B). Looking only to the evolution of a plane wave
one can calculate the interference fringes and also the positions 
of the center of the wave-packet, given by the derivative of the 
phase with respect to the momentum $p_0$. 

The output of the interferometer can be precisely calculated. 
We have to solve the Schr\"odinger equation: 
\begin{equation}
i\hbar \frac{\partial \ket\psi}{\partial t} = H(t)\ket\psi
\end{equation}
where the Hamiltonian $H(t)$ is the sum of the free Hamiltonian $H_0$ and
the Raman interaction $\VR(t)$ with
\begin{equation}
H_0 = \frac{\hat p^2}{2m} + \hbar\omega_{12} \ketbra{g_2}{g_2}
\end{equation}
where $\hbar \omega_{12}$ is the energy difference between the two states \ket{g_2} and \ket{g_1} and
\begin{equation}
\VR(t) =  
\frac{\Omega(t)}2 \left(e^{i\left(\phi_\mathrm R(t) + k_\mathrm{eff}\hat x\right)}
      \ketbra eg + h.c.\right)
\end{equation}
where $\Omega(t)$ is equal to the Rabi pulsation of the two photon Raman 
transition ($\Omega_0$ during the pulses and 0 elsewhere), $\hbar k_\mathrm{eff}$ 
is the effective recoil transferred by the Raman pulses and $\phi_\mathrm R(t)$ is 
the phase difference between the two lasers.

We choose to 
solve this Hamiltonian using the plane wave basis. We then have a two levels 
system $\ket{g_1, p_0}$ and $\ket{g_2, p_0 + \hbar k_\mathrm{eff}}$. We also write 
the phase of the laser as $\phi_\mathrm R(t) = \omega_{12}t + \phi(t)$ and change
the ket \ket{g_2} to $e^{i\omega_{12}t}\ket{g_2}$ (rotating frame). The Hamiltonian is then this 2x2 matrix:
\begin{equation}
H=\begin{bmatrix}
p_0^2/2m & \Omega(t)e^{i\phi(t)}/2 \\
\Omega(t)e^{-i\phi(t)}/2 & (p_0+\hbar k_\mathrm{eff})^2/2m
\end{bmatrix}
\end{equation}
This Hamiltonian contains the phase shift induced by the kinetic energy
and the phase transferred from light to atoms during a Raman transition. Assuming 
that $\phi(t)$ is almost constant during the pulse, we obtain that  
from \ket g to \ket e we add a phase $\phi(t)$ and from \ket e to \ket g we remove
this phase.

The final proportion of atoms transferred to state \ket e will vary as 
$\sin(\Phi)$ with $\Phi = \phi_A - \phi_B$ is the phase difference of the two paths with: 
\begin{eqnarray}
\phi_A & = & \phi_1 - \phi_3 + 
                  \phi^{\mathrm{sel}}_A(p_0) + \phi^{\mathrm{mes}}_A(p_0) \\
\phi_B & = & \phi_2 - \phi_4 
                  + \phi^{\mathrm{sel}}_B(p_0) + \phi^{\mathrm{mes}}_B(p_0) 
\end{eqnarray}
where $\phi_i = \phi(t_i)$ are the phases at the time $t_{1,2,3,4}$ of the four Raman pulses and $\phi^{\mathrm{sel/mes}}_{A/B}(p_0)$ are the phases due to kinetic energy: 
\begin{eqnarray}
\phi^{\mathrm{sel}}_A(p_0) &=&  \phi^{\mathrm{mes}}_B(p_0)  = 
                   \TRamsey \frac{(p_0 + \hbar\keff)^2}{2m\hbar} \\
\phi^{\mathrm{sel}}_B(p_0) &=&  \phi^{\mathrm{mes}}_A(p_0) = 
                   \TRamsey \frac{p_0^2}{2m\hbar} 
\end{eqnarray}

The phase can also be written as 
\begin{equation}
\Phi = \Phi_\mathrm{laser} + \Phi_\mathrm{kinetic}
\end{equation}
with $\Phi_\mathrm{laser} = (\phi_1 - \phi_3) - (\phi_2 - \phi_4)$
and $\Phi_\mathrm{kinetic} = 
(\phi^{\mathrm{sel}}_A(p_0) + \phi^{\mathrm{mes}}_A(p_0))
-(\phi^{\mathrm{sel}}_B(p_0) + \phi^{\mathrm{mes}}_B(p_0))$.

We observe that $\Phi_\mathrm{kinetic} = 0$ when the velocity of atoms
is constant between the pulses. In the case where there are uniform
forces that change $p_0$, we have to integrate the kinetic energy and
we obtain that:
\begin{equation}
\Phi_\mathrm{kinetic} = 
    \frac{k_\mathrm{eff}}m\int_0^\TRamsey (p_0(t_1+\tau) - p_0(t_3+\tau)) \mathrm d\tau
\end{equation}
The phase is a direct measurement of the momentum variation between the 
first pair of pulses and the second one.

Let us consider the case where there is a small change of velocity
$\delta v$ between the second and third pulse. The sensitivity of the
interferometer is proportional to the derivative of $\Phi$ with respect to
$\delta v$. 

Using the more general value of $\Phi$, and the fact that
$\phi^{\mathrm{sel}}_A(p_0) = \phi^{\mathrm{mes}}_B(p_0)$ and
$\phi^{\mathrm{sel}}_B(p_0) = \phi^{\mathrm{mes}}_A(p_0)$, we calculate that
\begin{equation}
\frac{\partial \Phi}{\partial \delta v} = m\left(
\frac{\partial\phi^{\mathrm{sel}}_A}{\partial p} - 
\frac{\partial\phi^{\mathrm{sel}}_B}{\partial p} \right)
\label{eq:sens}
\end{equation}

This equation tell us that in order to compute the sensitivity of 
the interferometer, we need only to calculate precisely the phase 
shift accumulated during half of the interferometer. Provided that 
the forces are uniform, this phase can be calculated for plane waves only. 

Further more, 
in quantum mechanics, the mean position $\left<X\right>$ of a wave-packet can be calculated
as the derivative of the phase with respect to momentum:
\begin{equation}
\left<X\right> = \hbar\frac{\partial\phi}{\partial p}
\label{eq:position}
\end{equation}
The equation \ref{eq:sens} shows directly that the sensitivity is proportional to
the spatial separation of the wave-packet:
\begin{equation}
\frac{\partial \Phi}{\partial \delta v} = \frac m\hbar \left(\left<X_A\right> - \left<X_B\right>\right) 
\label{eq:sensitivity}
\end{equation}

\section{Large area interferometer}

The interferometer based on large momentum transfer (LMT) Bloch pulses
is similar to the usual Ramsey Bord\'e interferometer.  Raman pulses
are used in the same configuration and the LMT Bloch pulses are simply
inserted between the two $\pi/2$ pulses used either for the selection or
the measurement of the velocity. A first implementation consists in
accelerating and then decelerating the atoms of one arm of the
interferometer for the selection (and then the atoms of the other arm
for the measurement). The temporal sequence of such an interferometer
is presented on Fig.~\ref{fig:LMTSimple}.

In the following section, we will present how this acceleration is implemented, and
then the numerical calculation of the phase shift $\Phi$ of 
such an interferometer.

\subsection{Implementation of LMTBS in the set-up}

\begin{figure}
\begin{center}
\includegraphics[width = .9\linewidth]{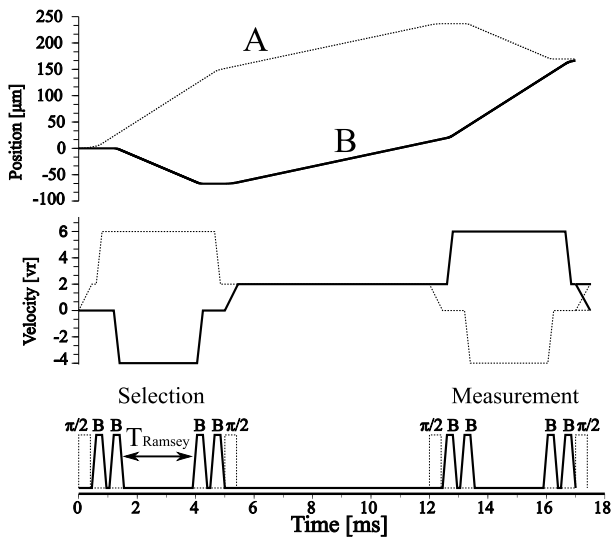}
\end{center}
\caption{Temporal sequence of the interferometer for N=2. From bottom to top: intensity of the Raman (dashed line)
and Bloch (solid line) beams; velocity of the atoms in the first and second arms of the interferometer;
trajectories of atoms in the two arms.}
\label{fig:sequence}
\end{figure}

In our experiment, another arrangement slightly more complicated than the one on Fig.~\ref{fig:LMTSimple} is used: we accelerate successively each arm of the interferometer
with two LMT pulses of opposite directions and then decelerate them with two other pulses. As we will see later, in such a way the contribution of
some systematic effects is reduced because the two arms are illuminated with Bloch beams in a more symmetric way.
The whole interferometer used in our experiment is presented on Fig.~\ref{fig:sequence}. It is realized with four Raman $\pi/2$ pulses and eight LMT Bloch
pulses (labeled B on the figure~\ref{fig:sequence}): each $\pi/2$ pulse is associated with two LMT pulses used to accelerate
selectively each arm of the interferometer.

After the first $\pi/2$ pulse, atoms are in a superposition of states
\ket{g_1,p_0} and \ket{g_2,p_0 + 2\hbar k}. A first LMT pulse is used
to accelerate atoms with initial momentum $p_0 + 2\hbar k$. Their
momentum became $p_0 + 2\hbar k + 2N\hbar k$ after $N$
oscillations. Then atoms with initial momentum $p_0$ are accelerated
in the opposite direction.  Therefore a superposition of
\ket{g_2,p_0+2\hbar k+2N\hbar k} and \ket{g_1,p_0 - 2N\hbar k} states
is created. After a delay \TRamsey, the same sequence of two LMT
pulses is repeated but in reverse order, to bring back atom states in
the initial superposition of \ket{g_1,p_0} and \ket{g_2,p_0 + 2\hbar
  k}. Finally, the sequence is ended with a second Raman
$\pi/2$-pulse. The wavefunction of atoms in the state \ket{g_2}
results from the interference between atoms transferred during the
first $\pi/2$ pulse and atoms transferred during the second one of the
selection sequence.

The interferometer is then closed in position space using a second set
of Raman and LMT pulses identical to the first one.  As for the
regular interferometer, the resulting signal of the interferometer
(population in \ket{g_1} and \ket{g_2}) vary as $\sin(\Phi)$, where
$\Phi$ is the phase difference between the two arms of the interferometer. 

On can write $\Phi$ as the sum of three phases:
\begin{equation}
\Phi = \Phi_\mathrm{Raman} + \Phi_\mathrm{Bloch} + \Phi_\mathrm{kinetic}
\end{equation}
The Raman and kinetic phase are the same as for the usual Ramsey-Bord\'e interferometer. 
The Bloch phase includes all the phases shift during the Bloch separation, due to
the phase of the laser, the light shift of atoms in the lattice and the kinetic energy. 

The Bloch phase is the difference between the Bloch phase $\Phi_\mathrm{Bloch}^{A/B}$ accumulated in the path
A and B. For a given sequence of the optical lattice (velocity and amplitude) and an initial momentum $p_0$, we calculate
the coefficients $c_l$ (c.f. eq.~\ref{eq:supermatrice}) corresponding to the evolution of the state \ket{p_0} (resp. \ket{p_0 + 2\hbar k}) using the model described earlier. The contribution of the first pulse to Bloch phase is deduced from the phases of the coefficients $c_{-N}$ (resp. $c_{N+1}$). 
The same method is used for all Bloch pulses. 
This allow us finally to calculate
the total phase for the 2 first beamsplitter, $\Phi^{A/B}_\mathrm{sel}$. From this
phase we deduce the position of the wave packets $\left<X_{A/B}\right> = \frac{\partial \Phi^{A/B}_\mathrm{sel}}{\partial p_0}$.
The sensitivity of the
interferometer can be deduced using equation (\ref{eq:sensitivity}).

\subsection{Experimental results}

The experimental setup and the main results have been described in a previous paper \cite{clade_PRL2009}. We briefly remind the
results to compare them with the theoretical model. We use rubidium 87 atoms, consequently  the two levels coupled by the Raman transition
are the \ket{F=2} and \ket{F=1} hyperfine level of the ground state.

\begin{figure}
\begin{center}
\includegraphics[width = .9\linewidth]{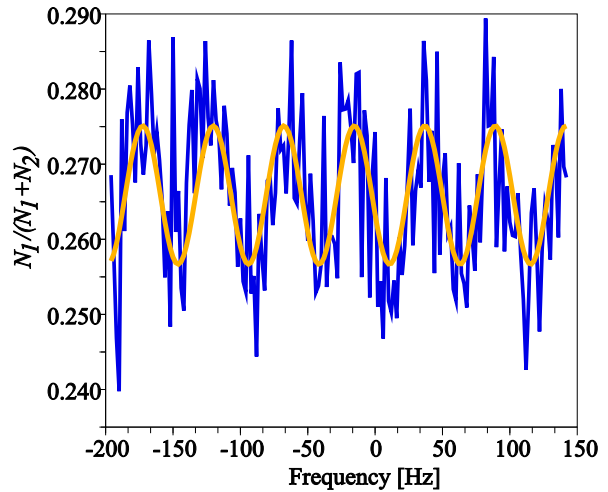}
\end{center}
\caption{Interference fringes observed using the scheme depicted on Fig.~\ref{fig:sequence}. On this figure, the main source of noise is the detection noise.}
\label{fig:Interference}
\end{figure}

The Fig.~\ref{fig:Interference} shows the proportion of atoms detected in the internal state $F=1$ as a function of the frequency
difference between selection and measurement using the time sequence reported on Fig.~\ref{fig:sequence}.

The periodicity of the fringes (Fig.~\ref{fig:Interference}) is about $\deltasep = 50\unit{Hz}$ for the time sequence
depicted on Fig.~\ref{fig:sequence}.

Quantitatively, $\deltasep$ can be compared with the periodicity of fringes of an interferometer without LMT Bloch pulses. With a
delay of 5\unit{ms} between the two Ramsey pulses of the selection, the periodicity would have been 200\unit{Hz}. Therefore with LMT pulses,
the sensitivity of the interferometer is improved by about four.

We can define an effective time $T_\mathrm{eff}= 1/\deltasep$, which is of about 20\unite{ms}. 
Because the fringes spacing depends only on the separation of the two arms of the interferometer, the effective time
$T_\mathrm{eff}$ can be viewed as the Ramsey time in a regular Ramsey-Bord\'e interferometer which leads to the same spatial
separation of the two wave packets.

\begin{figure}
\begin{center}
\includegraphics[width = .9\linewidth]{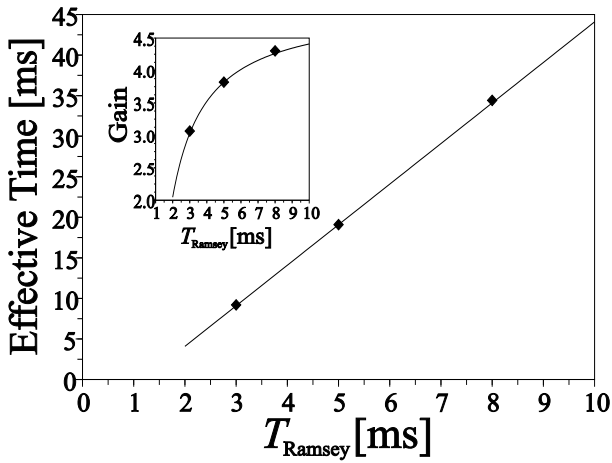}
\end{center}
\caption{Effective time $T_\mathrm{eff}$ as a function of the total time of the selection. The temporal sequence described Fig.~\ref{fig:sequence} is kept constant, except for \TRamsey. Solid curve: Numerical calculations. Inset: sensitivity gain of the LMT interferometer compared to a regular Ramsey Bord\'e interferometer of the same total duration.}
\label{fig:CompExpTh}
\end{figure}

More rigorously, $T_\mathrm{eff}$ and the spatial separation $\left<X_A\right> - \left<X_A\right>$
of the two wave packets are linked through the following equation:
\begin{equation}
T_\mathrm{eff} = \frac{\left<X_A\right> - \left<X_B\right>}{2v_r}.
\label{eq:TeffvsSeparation}
\end{equation}


The dependence of $T_\mathrm{eff}$ with the duration of the selection sequence has been also investigated experimentally. The
duration of the time-sequence of the interferometer is adjusted by changing \TRamsey\ keeping the same $\pi/2$-Bloch-Bloch sequence.
The variation of $T_\mathrm{eff}$ versus \TRamsey\ is
reported again on Fig.~\ref{fig:CompExpTh}. There is an excellent agreement between experiment and calculations.

In the inset of the Fig.~\ref{fig:CompExpTh}, we have also plotted the gain in resolution (ratio between
the effective time and the real time) as a function of the duration of the selection. At the limit where \TRamsey\ is long, the
duration of the pulses can be neglected and the gain will be exactly $2N+1$, i.e. 5 for our parameters. At shorter times, this
gain is smaller.

The experimental data fits well with the predicted time which is a strong evidence that the fringes observed
on Fig.~\ref{fig:Interference} are due to the LMT pulses.

\subsection{Light shift in the interferometer}
Light shift in the interferometer are one of the main source
of systematic effect. Furthermore they can induce a reduction 
of the contrast of interference.

The mean energy of atoms is indeed shifted when the lattice is
switched on. This shift is not the same between the different bands
and it depends on the intensity of the lattice. Our scheme (see
Fig.~\ref{fig:sequence}) is on average symmetric: that is the time
spent by atoms in a given energy band is equal for each arm of the
interferometer. Consequently, there is no associated systematic shift
if the intensity seen by atoms is constant. However, experimentally,
this condition is never realized because of temporal fluctuations of
the laser intensity and because of motion of the atoms through the
spatial profile of the laser beam. In the first case, fluctuations of
the intensity will be the same for each atom, leading to a phase noise
in the interference pattern.  In the second case, the light shift
contribution will be averaged over all the atoms, leading to a
reduction of the fringes contrast.

The light shift are implicitly included in the phase $\Phi_\mathrm{Bloch}$ obtained 
by integrating the Schr\"odinger equation. This phase depends on the intensity of the laser over time. 
For small variation of intensity, one could linearized the variation of the phase and
compute the functional derivative of the phase with respect to the intensity:
\begin{equation}
a(t) = \frac{\delta\Phi}{\delta I(t)}
\end{equation} 

In the experiment the
intensity of the laser is modulate by an acousto-optic modulator. The intensity
is therefore the product of the modulation ($M(t)$) due to the AOM and the intensity without modulation 
$I_0(t)$. If we assume that
the modulation for the LMT pulse does not fluctuate ($\delta M(t) = 0$), then the fluctuation of the phase 
is mainly due to fluctuations of $I_0(t)$ around its constant mean value $I_0$. 

In the following instead of calculating $\frac{\delta\Phi}{\delta I(t)}$, we choose to calculate the functional
derivative with respect to the normalized derivative of $I_0(t)$, $1/I_0\, \mathrm dI_0(t)/\mathrm dt$, that we denote by $g_I(t)$.  
We choose to use the derivative, because it is more convenient numerically and also lead
to a dimensionless functional derivative.
At first order, one can therefore write that: 
\begin{equation}
\delta\Phi = \int_{-\infty}^\infty g_I(t)\frac1{I_0}\frac{\mathrm dI_0}{\mathrm dt}dt\ .
\end{equation}
In the following, $g_I$ is referred as the sensitivity function to the relative derivative of intensity. 

\begin{figure}
\begin{center}
\includegraphics[width = .9\linewidth]{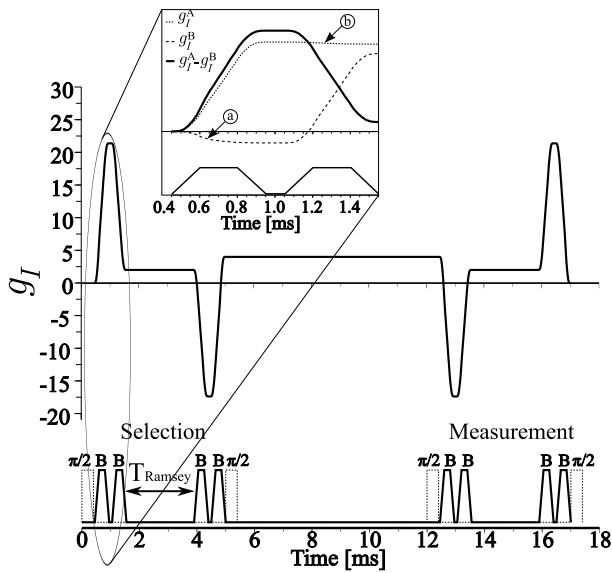}
\end{center}
\caption{Sensitivity function of the interferometer to the relative derivative of intensity. The temporal sequence is described
Fig.~\ref{fig:sequence}. Inset: zoom on the two first LMT pulses. The contribution from the two arms of the interferometer
are displayed separately. A constant phase shift corresponding to the average light shift of an atom in the lattice is subtracted
from both curves.}
\label{fig:Gint}
\end{figure}

Numerically, $g_I(t)$ is computed by looking to the variation 
of the phase $\Phi_\mathrm{Bloch}$ for a small change of the intensity
$I_0(t^\prime)$ for $t^\prime > t$ (Heaviside function). The temporal evolution of the function $g_I$ during the sequence
described previously on Fig.~\ref{fig:sequence} is reported on
Fig.~\ref{fig:Gint}.

The inset is a zoom of $g_I$ at the beginning of the sequence. In the
inset, the contributions to $g_I$ of each arm ($g_I^A$ and $g_I^B$)
are plotted. In the experiment, atoms that are in excited band experience 
on average, the mean potential of the lattice, $U_0/2$, whereas atoms in the 
first band, are in dark region (blue detuned lattice) and less subjected to light shift. 

The phases in the inset of Fig.~\ref{fig:Gint} are calculated with eq.~\ref{eq:Hamilt2} in
which the common light shift $U_0(t)/2$ is left out. This changes the values of $g_I^A$ and $g_I^B$ but 
not the value of $g_I$. 
Within this energy reference, an atom moving fast with respect to the lattice does
not see a phase shift ($\left< \cos(2k\hat x)\right> =0$) whereas an
atom trapped into the lattice sees a phase shift ($\left< \cos(2k\hat
x)\right> \approx -1$). Therefore the main contribution to the phase
shift originates from the lower band.  Consequently, it explains why
the initial increase of $g_I$ principally arises from the first band.

This phase shift is almost compensated by the other arm during the
second LMT pulse. Indeed, the first two LMT pulses do not act in a
symmetric way between the two arms because non-accelerated atoms are
not in the same excited band during the first and second LMT
pulses. During the first LMT pulse, non-accelerated atoms endure a
small light shift (see the variation of $g_I^B$, around t=0.6\unit{ms}
in the inset of Fig.~\ref{fig:Gint}, arrow labeled a). While during
the second LMT pulse, non-accelerated atoms are moving too fast with
respect to the lattice to see a light shift (see the plot of $g_I^A$
which remains constant during the second LMT, at $t$ around
1.2\unit{ms}, arrow labeled b).

In the case of the interferometer with only four LMT pulses (see
Fig.~\ref{fig:LMTSimple}), the initial phase due to the first LMT
pulse is compensated only at the end of the interferometer, after
about 10\unit{ms}, whereas in our scheme, 90\% of this phase shift is
compensate after 1\unit{ms}. The sensitivity of our interferometer to
a linear variation of the phase ($\int g_I(t) \mathrm dt$) is then
reduced by a factor of 10 compared to an interferometer with four LMT
pulses.

\def\dad#1#2{\frac{\mathrm d #1}{\mathrm d#2}} From the sensitivity
function $g_I$, the reduction of the contrast due to the motion of atoms
during the interferometer can be precisely calculated. For that, the
phase shift accumulated along a trajectory must be calculated. At
first order, it is expressed as
\begin{equation}
\phi = \int g_I \frac1{I_0}\dad {I_0}t \mathrm dt \approx \left(\int g_I
\mathrm dt \right) \frac1{I_0}\left.\dad {I_0}t\right|_{t=0 }
\end{equation}
The contrast $C$ is given by $C=e^{-\Delta\phi^2/2}$ where
$\Delta\phi$ is the variance of the phase $\phi$. $\Delta\phi$ is
given by:
\begin{equation}
\Delta\phi=\sigma_{\dot{I}} \left|\int g_I \mathrm dt \right|,
\end{equation}
where $\sigma_{\dot{I}}$,  which represents the relative variance of $\dad{I_0}{t}$ over the entire atomic cloud, is calculated with
\begin{equation}
\sigma_{\dot{I}} = \frac1{I_0} \sqrt{\left<\left(\dad{I_0}{t}\right)^2\right>}.
\end{equation}

To evaluate $\sigma_{\dot{I}}$ numerically, Gaussian distributions are
used to described the position and the velocity of the atoms.  Then
$\sigma_{\dot{I}}$ can be expressed as:
\begin{equation}
\sigma_{\dot{I}} = 4\sqrt{2} \sigma_v \frac{\sigma}{w_L^2} \frac1{1+8\frac{\sigma^2}{w_L^2}} .
\end{equation}
in which $\sigma$ denotes the RMS value of the position of the atoms
in one direction and $\sigma_{v}$ the RMS value of their velocity,
$w_L$ represents the waist of the Gaussian laser beam. With ours
experimental parameters ($w_L = 2\unit{mm}$, $\sigma = 600\unit{\mu
  m}$, $\sigma_v = 2v_r$), the value of $\sigma_{\dot I}$ is around
$5.9 \unit{s^{-1}}$.

For the eight LMT pulses scheme used in our experiment, we have
calculated that: $\int g_I \mathrm dt \approx 48\times 10^{-3}
\unit{rad\cdot s}$.  Therefore, the values of $\Delta\phi$ and $C$ are
: $\Delta\phi =0.28 \unite{rad} $ and $C=96 \%$.

With the four pulses scheme, those values become $\Delta\phi = 2.8 \unite{rad} $ and $C=2\%$.
This clearly shows that in our experiment, the eight LMT pulses scheme must be used to observe interferences.

We emphasize that there is a correlation between the velocity and the position of atoms because of the small time of flight
before the beginning of the interferometer. This results in a biased intensity variation leading to a systematic effect in
the interferometer. This effect can be canceled by inverting the order of the two LMT pulses for each beamsplitter
because the sign of $g_I$ and consequently the one of $\phi$ is reversed.


\section{Improving LMT beamsplitter}

\subsection{Current limitations}

The contrast of the fringes observed Fig.~\ref{fig:Interference}
is ten times smaller than the one without LMT pulses. 
This weak contrast can not be explain with light shifts.
In the experiment, light shift is under control when the LMT pulses are used on both arms of the interferometer.
Other sources contribute to this reduction. The main one arises from inhomogeneities in the initial momentum distribution of atoms
and intensity of the laser. The maximum efficiency computed has been obtained for an
atom with a defined velocity and a given laser intensity. But in our set-up, the initial velocity distribution is spread over
about $0.5\,v_r$, consequently the effective efficiency for a LMT pulse is reduced from $98\%$ to $95\%$.
Moreover, because of the initial size of the cloud and the finite waist of the laser, the variations of the intensity seen by atoms
are in the order of $20\%$. As it can be easily see on Fig.~\ref{fig:probaU0}, this effect pushes away the efficiency from the optimal
configuration by few percents. The contribution of spontaneous emission of atoms trapped in the lattice is also non negligible.
In our regular experiments with Bloch oscillations~\cite{clade:052109}, atoms, located at the bottom of the lattice wells, are oscillating in the
first band. As the potential of the lattice is blue detuned, the spontaneous emission is reduced, because the light intensity
is minimal at the position of atoms~\cite{clade:052109}. Unfortunately, this demonstration does not apply to atoms in the excited
bands. The estimated spontaneous emission rates of those atoms is around $4\%$. 

All these effects added together reduce
significantly the efficiency of each of the 8 LMT beamsplitter used in the interferometer and explain the reduction
of contrast that we have observed. 
However, using a laser beam with a larger
waist, a higher intensity (and a larger detuning) should allow us to reduce significantly these effects.

\subsection{Sequence of a 10 recoil beamsplitter}

As mention previously, it is not possible to realize a larger momentum separation while keeping a good efficiency with the
scheme of the LMT, in which constant acceleration is used (c.f. Fig~\ref{fig:etavsT}).

To overcome this limitation, a process with different acceleration steps can be foreseen. The first step is identical to the one
described earlier, with two Bloch oscillations. For the second one, the depth of the lattice is raised to enable a larger
acceleration.

As explained previously, a long time and a relatively weak lattice are necessary to realize the first splitting with
a good efficiency. For a given time, the optimal value of the amplitude of the lattice is a compromise between good
adiabaticity of Bloch oscillations in the first band and a high Landau-Zener tunneling for the third band ($n=3$). However, this
condition is less restricting when atoms are in a higher band ($n>3$). For those atoms, the Landau-Zener tunneling is higher
so one can expect a better efficiency with the same duration or the same efficiency with a shorter duration. That is, a larger
acceleration can be use to increase separation with a good efficiency.

\begin{figure}
\begin{center}
\includegraphics[width = .9\linewidth]{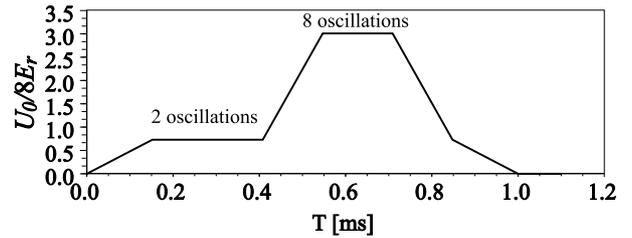}
\end{center}
\caption{Scheme of a beamsplitter with 10 oscillations.}
\label{fig:10oscillations}
\end{figure}

The realization of a beam splitter with ten oscillations is presented on Fig.~\ref{fig:10oscillations}. The initial separation with
two BOs is achieved in 130\unite{\mu s} while the second separation with eight BO is done in only 160\unite{\mu s}.
The maximal efficiency is calculated to be about $98\%$. This is very close to the efficiency with two BOs
beamsplitter. The main losses arise from the initial splitting of two recoils. The additional losses come from the
lack of adiabaticity during the change of the lattice intensity. As Bloch oscillations is a very efficient process, no significant
losses arises from the additional recoils transferred after the first two oscillations. The main limitations will come from the reduction of contrast due to
intensity fluctuations.

\begin{figure}
\begin{center}
\includegraphics[width = .9\linewidth]{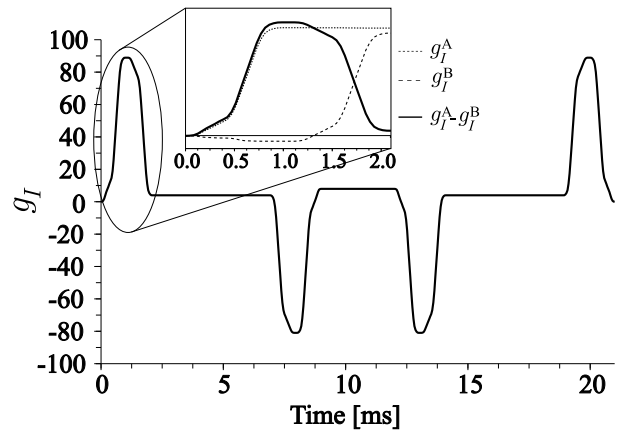}
\end{center}
\caption{Sensitivity function $g_I$ of the interferometer using a 10 Bloch oscillations LMT pulse, as described in Fig.~\ref{fig:10oscillations}.}
\label{fig:Sensitivity_10}
\end{figure}

To estimate this limitation, the sensitivity function $g_I$, with ten BOs, is plotted on
Fig.~\ref{fig:Sensitivity_10}. It is significantly larger than the one with two BOs because of the increase of the depth of the
lattice. The sensitivity to a linear variation of intensity $\int g_I \mathrm dt$ is about $95 \times 10^{-3} \unit{rad\cdot s}$. 
That is twice as much as the one for the interferometer with two BO (see section \ref{section2}). This is the reason why the improved LMT
was not implemented yet on our current apparatus (the visibility of fringes for the "two BOs scheme" is already quite weak). 

%

\section{Conclusion}

In this paper, we have theoretically and experimentally studied the implementation of a large momentum beamsplitter in a
Ramsey-Bord\'e interferometer. We have realized a separation of 10 recoils between the two arms of the
interferometer. By reducing the contrast of the fringes, light shift is the main limitation of this
interferometer. Nevertheless, we have developed a method to reduce this effect by a factor of ten and so realize
an interferometer with a separation of ten recoils between the two arms. Our model is in very good agreement with the experiment,
in which a gain of four in the resolution, compared to a usual interferometer, was observed. This method seems to be very promising
for the realization of an interferometer with a separation of several tens of recoil velocities.

The limitation due to light shift has been strongly reduced by accelerating successively both arms of the interferometer. Moreover a
way to 
systematically cancel it would be to accelerate both arms of the interferometer simultaneously (instead of doing it successively).
This can be done by applying two counter-propagating accelerated lattices. This scheme is already used by the
Stanford/Berkeley group \cite{muller2009_PRL240403}. In their experiment, a high order Bragg beamsplitter is initially used,
before the double BO acceleration. We are also currently investigating this possibility of using the double BO acceleration but with an
initial velocity separation given by a two photon Raman pulse. It seems that only two recoils is two small to perform a good separation and a higher initial separation may be required. A relevant method could be to use the double-diffraction technique described in Ref.~\cite{leveque:2009} which allow an initial separation of 4 recoils. 
In this scheme, light shift would be completely suppressed.
One of the remaining source of noise will be then the phase/vibration noise. This noise can be calculated with a
numerical model identical to the one used in this paper.

\section{Acknowledgements}

This work is supported in part by IFRAF (Institut Francilien de Recherches sur les Atomes Froids), and by the Agence Nationale pour la Recherche, FISCOM Project-(ANR-06-BLAN-0192).



\end{document}